# π-π Interaction-facilitated formation of interwoven trimeric cage-catenanes with topological chirality

Lihua Chen,[1] Zhenghong Chen,[1] Weihao Wang,[1] Chenhao Chen,[1] Kuboi Yoshiaki,[1] Chi Zhang,[1] Chenfei Li,[1] Shaodong Zhang[1],*

[1]School of Chemistry and Chemical Engineering, Shanghai Jiao Tong University, Shanghai 200240, China

## Abstract

Catenanes as interlocked molecules with a nonplanar graph have gained increasing attention for their unique features such as topological chirality. To date, the majority of research in this field has been focusing on catenanes comprising monocyclic rings. Due to the lack of rational synthetic strategy, catenanes of cage-like monomers are hardly accessible. Here we report on the construction of an interwoven trimeric catenane that is composed of achiral organic cages, which exhibits topological chirality. Our rational design begins with a pure mathematical analysis, revealing that the formation probability of the interwoven trimeric catenane surpasses that of its chain-like analogue by 20%; while driven by efficient template effect provided by strong π-π stacking of aromatic panels, the interwoven structure emerges as the dominant species, almost ruling out the formation of the chain-like isomer. Its topological chirality is unambiguously unravelled by chiral-HPLC, CD spectroscopy and X-ray diffraction. Our probability analysis-aided rational design strategy would pave a new venue for the efficient synthesis of topologically sophisticated structures in one pot.

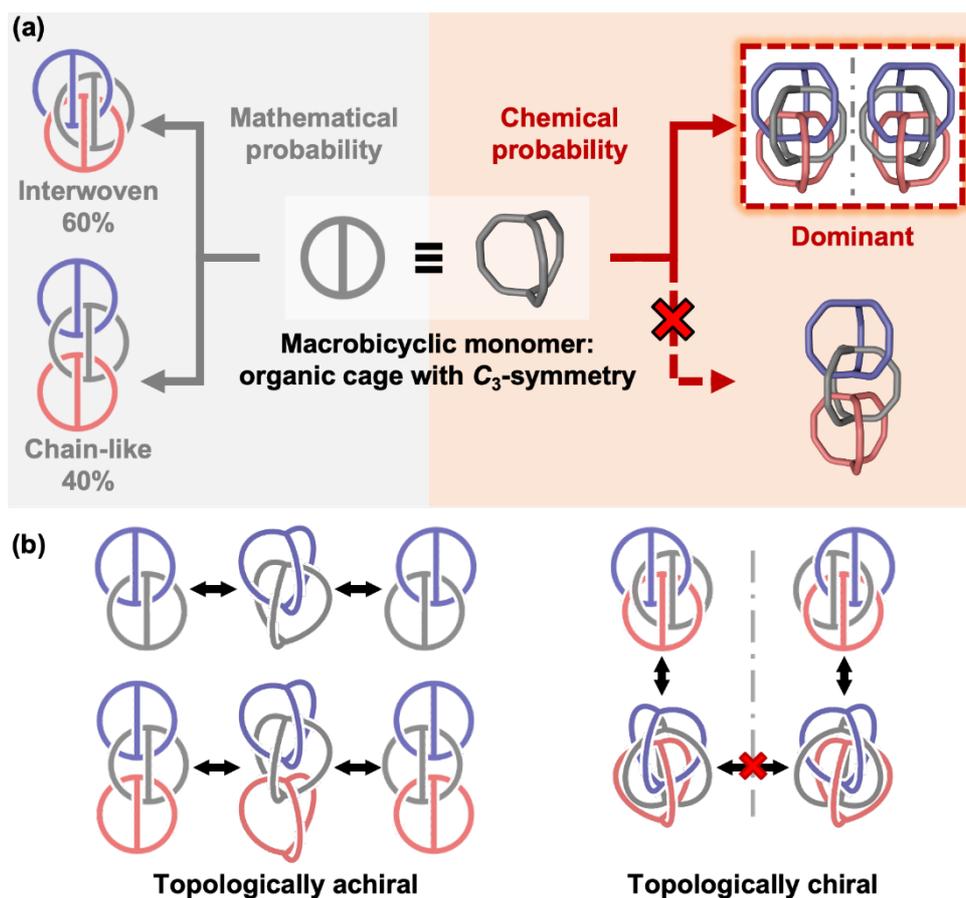

**Fig. 1 The formation of interwoven trimeric catenane with topological chirality as the dominant species**. **a**, Compared to its chain-like analogue, the interwoven trimeric catenane is statistically favoured and can evolve as a dominant chemical entity. **b**, Catenation of bicyclic monomers forms topologically achiral dimeric catenane and chain-like trimeric catenane, while it yields topologically chiral interwoven trimeric catenane.

Catenanes are a class of interlocked molecules characterised by a nonplanar embedded graph. These intricate structures have garnered significant interest in recent years, primarily due to their intriguing topological features and potential applications in molecular machines[1], catalysis[2,3], and other areas. Since the seminal work of Sauvage et al.[4], which utilised metal-ligand templation to enhance the preorganization of reactive components, the pursuit of synthesizing catenanes has become a captivating endeavour, challenging the boundaries of human imagination and synthetic proficiency[5–9]. To date, the majority of research in this field has focused on catenanes comprising monocyclic rings[9]. Notably, significant achievements have been made in constructing various molecular structures with the graph of prime links[10]. Among them, an isomeric pair of Borromean ring[11–12] and interwoven [3]catenane[13–15] are particularly appealing. The former (topology $6_2^3$ in Alexander–Briggs notation[16]) is interlocked such that all three rings fall apart upon fission of one ring, while the latter (topology $6_3^3$) still retains two catenated rings when one ring is cleaved. Catenanes have also evolved to incorporate more complex multi-annulated constituents[17–24], *i.e.*, three-dimensional cage-like monomers. On the other hand, although the synthesis of interpenetrated cages was first reported by Fujita and co-workers in 1999[17], due to the lack of rational design strategy, the only accessible structure of this kind is dimeric catenanes, which are either composed of coordination[17,21,22,25–27] or purely organic cages[18–20,24,28–30]. The trimeric cage-catenane was not achieved until the very recent serendipitous discovery by Mastalerz et al.[31], who elegantly synthesised the trimeric catenane composed of cube-shaped cages, guided by weak dispersion interactions and solvophobic effects. However, they failed to resolve this intricate molecular structure by X-ray diffraction. Besides, to the best of our knowlesge, the topology related properties of these catenanes have not been explored.

We herein report on the rational design and efficient synthesis of a trimeric cage-catenane, denoted as **TCC-1**, which is composed of three identical organic cages[28–30]. Inspired by Fujita's seminal works on aromatic stacking-mediated cage interpenetration[25–27], we selected a series of aromatic panels, by assuming their π-π stacking would facilitate their preorganisation for subsequent cage catenation. As the catenation process would yield a pair of topological isomers, *i.e.*, interwoven and chain-like trimeric cage-catenanes (Fig. **1**), the selective formation of one isomer presents a formidable challenge. We therefore started the rational design with a pure statistical analysis by comparing the formation probabilities of the two isomers (Fig. **1a**). Without considering the chemical circumstance, it reveals that the interwoven species outperforms its chain-like counterpart, exhibiting a formation probability of 60% compared to 40%. We further built up a refined statistical model by considering the probability density calculated under chemical circumstance, specifically accounting for the π-π stacking as a template effect. It shows that the interwoven structure becomes the dominating interlocked species.

Guided by this rationale, we employed dynamic condensation between trialdehyde aromatic panels and diamine linkers of suitable length in stoichiometry, leading to the formation of interwoven **TCC-1** as *the only interlocked species*, along with the corresponding cage monomer as a minor product. With the combination of chiral-HPLC, circular dichroism (CD) spectroscopy and single-crystal X-ray diffraction (SC-XRD), we unambiguously revealed the inherent topological chirality of the interwoven trimeric catenane, in contrast to the dimeric and chain-like trimeric analogues that can be converted between mirror images without bond cleavage and are thus topologically achiral (Fig. **1b**). Our investigation compellingly underscores that chirality can indeed be a topologically related characteristic, contingent upon both the number of monomers and the manner in which they are interconnected.

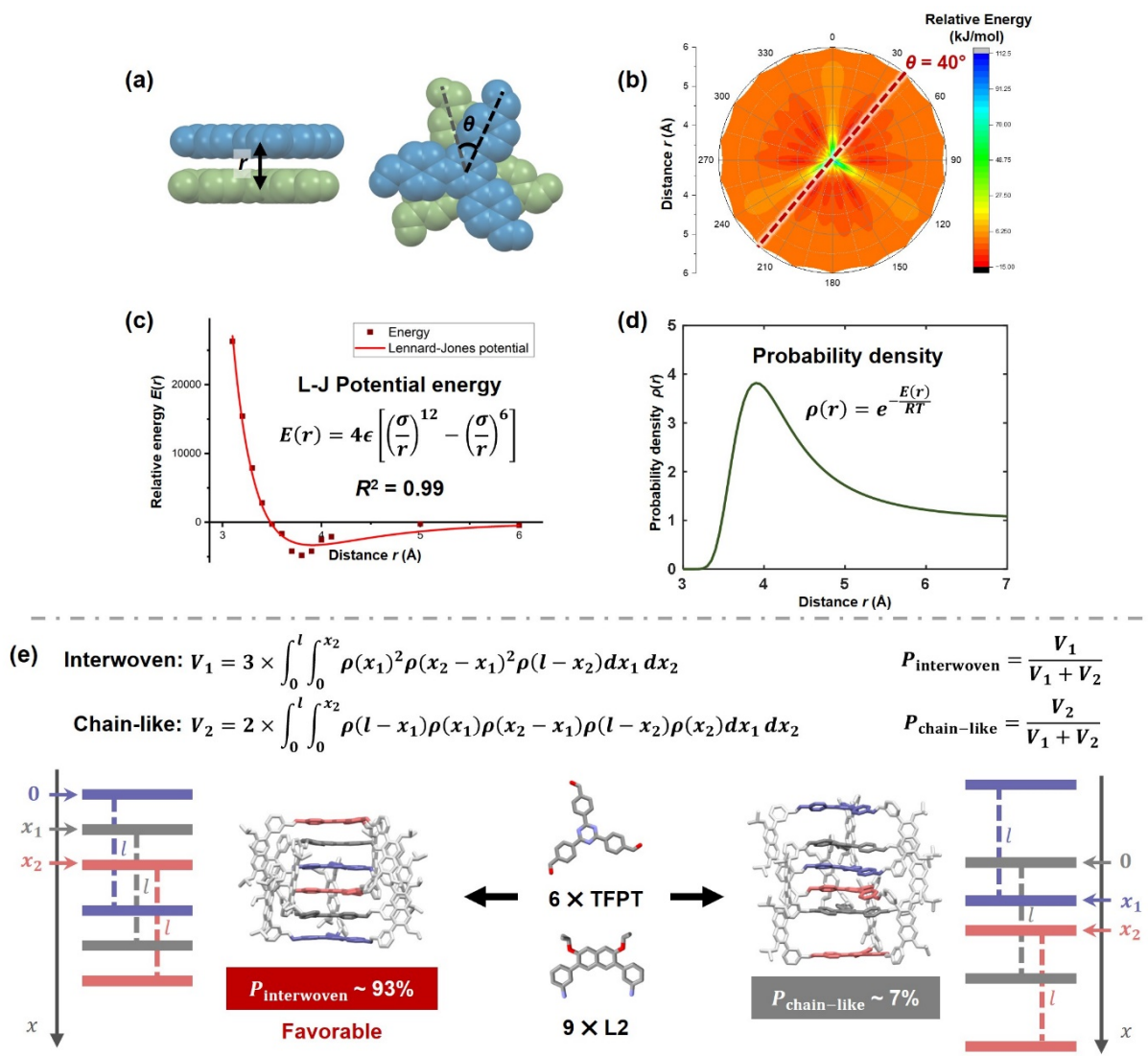

**Fig. 2 Comparison of formation probability of interwoven trimeric cage-catenane TCC-1 and its chain-like analogue**. **a**, Schematic representation of vertical stacking of two 1,3,5-tris(*p*-formylphenyl)triazine (**TFPT**) panels with a distance *r* and rotation angle *θ*. **b**, Relative energy between two **TFPT** panels as a function of *r* and *θ*. The relative energy is minimal when *θ* = 40º, and this angle is therefore chosen for the calculation of Lennard-Jones (L-J) potential *E*(*r*) as a function of distance *r* in **c**. **d**, Probability density *ρ*(*r*) of distribution of a **TFPT** to the most adjacent one as a function of distance *r*. **e**, The calculation of relative formation probability of interwoven and chain-like trimeric catenane, showing the interwoven species is dramatically dominating when the π-π stacking between **TFPT** panels is taken into consideration. The calculation is conducted with the constraint of fixed vertical distance *l* of two panels that belong to the same cage monomer.

**Results and discussion**

**Rational design assisted by mathematical analysis**

We recently have been focusing on the synthesis of dimeric catenane composed of organic cage monomers[28–30]. As a continuous endeavour, we sought to construct the trimeric catenane, which, *a priori*, includes two topological isomers of interwoven and chain-like structures (Fig. 1). We first set out to probe the formation probabilities of the two topological species with a pure mathematical model. This probability model based on phase spaces of degree of freedom was developed by simply considering the preorganization of six tritopic panels, which is a necessary and sufficient condition for the formation of catenated trimers (Supplementary Fig. 1). The step can be mathematically presented by the process where two panels (needed to form a cage monomer) penetrate four other preorganized panels (required to construct a catenated dimer), which shows that the formation probability of the interwoven trimer is 60%, slightly favoured to its chain-like counterpart with formation probability of 40% (Fig. **1a** and Supplementary Fig. 1, see detailed calculation in the Supplementary Information).

As previously proved by Fujia[25–27], Cooper[20] and us[28,29], the template effect provided by directional non-covalent interactions of the synthons is crucial for the formation of cage-catenanes. We therefore built up a refined statistical model with the probability density calculated by considering π-π stacking of 1,3,5-tris(*p*-formylphenyl)triazine (**TFPT**), namely the panel used for constructing the cage-catenane in the current study.

We first scanned the potential surface of two **TFPT** panels with the optimized geometry, which are aligned in parallel with a specific distance $r$ and rotation angle $\theta$ (Fig. **2a** and Supplementary Fig. 3). Their relative potential is then determined by varying the values of $r$ and $\theta$, as depicted in Fig. **2b**. The resulting contour plot exhibits higher energy levels when the two panels are nearly overlapping in the top view, owing to the strong repulsion in this conformation. However, the relative energy reaches the minimum at a rotation angle of approximately 40º. The energy $E(r)$ calculated with this angle is further fitted using the classical Lennard-Jones (L-J) potential, as illustrated in Fig. **2c**. With the determination of the potential function $E(r)$, the probability for a panel to take place at a certain distance from its adjacent panel should follow the density distribution function $\rho(r)$, which is induced from Boltzmann distribution at statistical equilibrium (Fig. **2d**).

As mentioned above, the formation of the trimeric catenated cage can be treated as a simple geometrical model consisting of six stacking panels (Fig. **2e**). Thus, the six panels will form five intervals, and for each interval the density distribution function $\rho(r)$ is only decided by the relative distance of two adjacent panels. Different arrangements of these parallel panels can lead to either interwoven or chain-like structures. For a specific arrangement order, each panel can still undergo slight movement under the constraint of π-π interaction, and the expression of $\rho(r)dl$ therefore represents the density of each infinitesimal length. Therefore, the sum of these densities in all arrangements ($V_1$ and $V_2$) is used to calculate the relative formation probability of interwoven and chain-like trimeric catenanes. Notably, the movements of the stacking panels occur at a significantly faster rate compared to the condensation reaction that links them to form the cage entity, which validates the assumption of Boltzmann radial distribution at equilibrium.

We employed a conventional cut-off scheme for non-adjacent interactions, which means that a panel's influence is limited to the space between its neighbouring panels and does not extend to the others. Hence, the overall density distribution function can be expressed as the product of the distribution functions in five distinct intervals. The expressions for the sum $V_1$ and $V_2$ are shown in Fig. **2e**, suggesting the cases of interwoven and chain-like structure, respectively. Therefore, the proportion of $V_1$ reflects the relative probability for forming the interwoven structure. Through this integration, the formation probability of trimeric interwoven catenane ($P_{\text{interwoven}}$) is about 93%, indicating the predominance of such species.

**Synthesis and characterization of interwoven trimeric cage-catenane**

Guided by the dominating formation probability of interwoven **TCC-1** determined in Fig. **2**, we first selected the tritopic panel **TFPT**, and a diamine linker, **L2**, of appropriate length so that the resulting monomeric cage is capable of accommodating two additional panels (Fig. **3**). Besides, we introduced lateral isopropyl groups on each linker to enhance the solubility and aid in the identification of reaction species using NMR spectroscopy (Fig. **3a,b**, details below).

The reaction between **L2** and **TFPT** was performed in chloroform with TFA as a catalyst (Fig. **3a**). Analysis of the crude mixture using matrix-assisted laser desorption/ionization time-of-flight (MALDI-TOF) mass spectrometry revealed a prominent ion peak at *m/z* 5875.97, indicating the formation of [9+6] protonated products, and another peak at *m/z* 1959.81 corresponding to [3+2] protonated products resulting from cycloimination between the reactants (Supplementary Fig. 75). These two species were later identified as the interwoven trimeric catenane **TCC-1** and monomeric cage **MC-6**, respectively (*vide infra*).

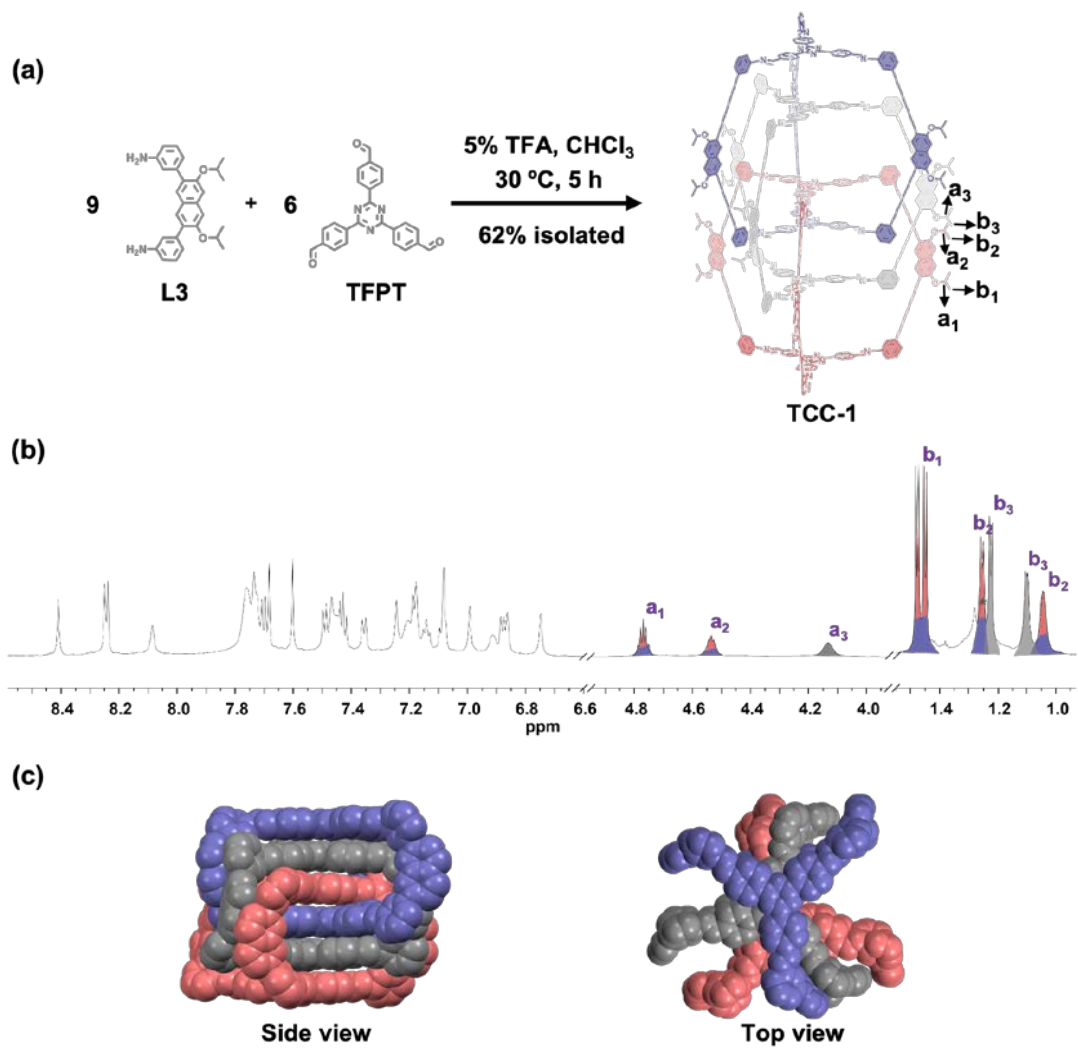

**Fig. 3 Synthesis and structural characterisation of interwoven trimeric cage-catenane TCC-1. a**, Schematic representation of reaction between diamine linker **L2** (3 equiv.) and trialdehyde panel **TFPT** (2 equiv.) to yield **TCC-1**. **b**, Truncated $^1$H NMR spectrum (500 MHz, CD$_2$Cl$_2$, 298.15 K) of pure **TCC-1**. Only characteristic chemical resonances are assigned for clarity, and full assignment is shown in Supplementary Information. **c**, The side and top views of crystal structure of **TCC-1** ambiguously show its interwoven feature.

**TCC-1** (62% isolated yield) was successfully purified using recycling semi-preparative gel permeation chromatography with dichloromethane as eluent (Supplementary Fig. 21). In its $^1$H NMR spectrum (Fig. 3b), three sets of protons with equal intensity are clearly observed, in contrast to the readily assignable protons of cage monomer (Supplementary Fig. 59). The assignments of the protons were aided by a combination of various 2D NMR spectroscopies (Supplementary Fig. 66–74). As compared to isopropyl protons on the outmost cage ($a_1$/$b_1$), significant upfield shifts of the corresponding protons on the interior cages ($a_2$/$b_2$ and $a_3$/$b_3$) are noticed, which can be attributed to the shielding environment withing the interwoven structure.

Single crystals of **TCC-1** suitable for X-ray diffraction were obtained by the gradual vapor diffusion of methanol into a chlorobenzene solution. **TCC-1** crystallised in the triclinic space group *P*1. The crystal structure analysis reveals its interwoven nature (Fig. 3c). It shows that three monomeric cages are triply interlocked, such that every window of a cage is intercalated by two twisted linkers, each belonging to a compositional cage. Efficient π-π stacking is achieved through the parallel alignment of six **TFPT** panels, with centroid distances of approximately 3.6 Å (Supplementary Fig. 12). These interactions are likely to serve as driving forces for preorganizing the reactants or intermediates into favourable positions for subsequent catenation.

As previously mentioned, the theoretical formation probability of the chain-like **TCC** is not insignificant (Fig. 1a and 2e). However, we were unable to experimentally obtain this species. We therefore employed diffusion-ordered NMR spectroscopy (DOSY) to investigate the species present in the reaction mixture (Supplementary Fig. 6). The DOSY NMR analysis of the reaction mixture in $CD_2Cl_2$ at 295 K revealed only one set of meaningful signals, with a diffusion coefficient ($D$) of $3.04 \times 10^{-10}$ $m^2 \cdot s^{-1}$ that corresponds to a solvodynamic radius ($R_s$) of 1.71 nm (Supplementary Fig. 5). It can be assigned to the interwoven **TCC-1**, identified by comparison with the pure compound (Supplementary Fig. 6–8), which is in line with its dimension of 1.75 nm determined by its crystal structure (Supplementary Fig. 12). From the DOSY analysis, it confirms that this species is not the chain-like structure, as it would exhibit a larger solvodynamic radius[31].

Our experiments thus validate our rational design strategy based on statistical analysis by considering the template effect provided by **strong π-π stacking,** which predicts a significant preference for the formation of the interwoven trimeric cage-catenane over the chain-like isomers. It is also worth mentioning that while our current study did not identify chain-like trimeric catenanes, their absence is not definitive. It is possible that their formation may be influenced by factors such as concentration, solvent properties, and other kinetic variables.

**Structural impact of linkers and panels on the formation of catenated species**

We further examined the impact of structural features of the precursors, namely linkers and panels, on the product distribution of monomeric and catenated cages (Fig. 4). Particularly, length and steric effect of the linkers, and π-π stacking strength of the panels were studied.

We started the investigation by using a model reaction with linker **L1**, which has a restricted length suitable for the formation of only dimeric catenanes as interlocked species (Table in Fig. 4). The reaction of **L1** with 1,3,5-triformylbenzene (**TFB**) yielded a mixture comprising monomeric cage **MC-1** and dimeric catenane **DCC-1**, with a conversion ratio of 56% and 21% at equilibrium, respectively. In contrast, when **L1** reacted with the larger-sized planar panel **TFPT**[32], the formation of predominantly catenated product **DCC-2** (in conversion of 83%) was observed, with minuscule amount of monomeric cage only detectable by MALDI-TOF (Supplementary Fig. 81), but not by [1]H NMR (Supplementary Fig. 17). Along with our previous studies[28,29], this comparison solidifies the evidence that the enhanced π-π stacking of the panel exerts a stronger driving force for the process of catenation.

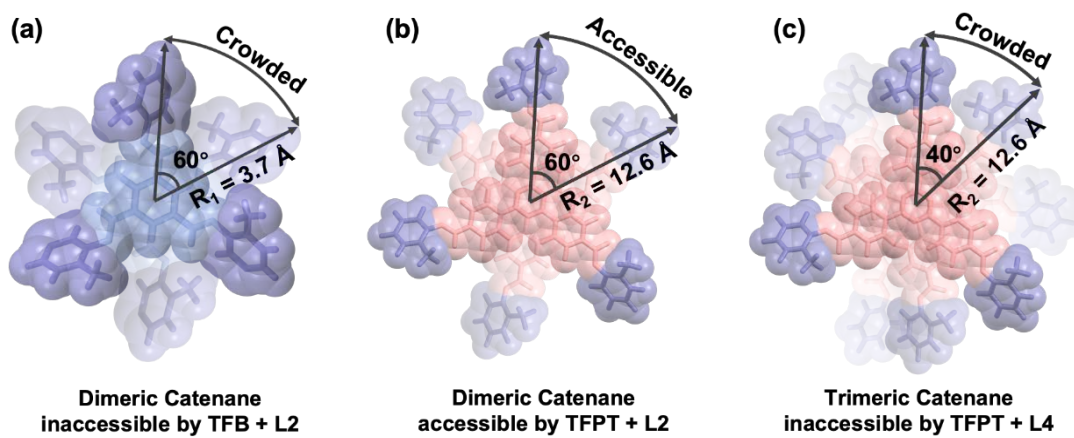

**Fig. 4 Structural impact of linkers and panels on the formation of catenated species**. The table summarises the product distribution of cage species, including monomeric cage (**MC**), and dimeric cage-catenane (**DCC**), and trimeric cage-catenane (**TCC**). **a,b,c** demonstrate the steric effect on the formation of cage-catenane.

The impact of π-π stacking was more pronounced when the longer linker **L3** was used in the reactions (Table in Fig. 4). When it reacted with the smaller panel **TFB**, only monomeric cage **MC-3** was produced (with a conversion of 35%). The lack of any interlocked species in the reaction with **TFB** suggests its strength of π-π stacking is insufficient to form stable enthalpic intermediate, *i.e.*, the host-guest complex formed by encapsulating **TFB** in the cavity of monomeric cage that is entropically unfavoured[29]. On the other hand, the reaction with the larger panel **TFPT** yielded a mixture of monomeric cage **MC-6** and trimeric catenane **TCC-1**, with conversion ratios of 4% and 72% at equilibrium, respectively.

The steric hindrance also exerts a notable influence on the catenation, as demonstrated by the reactions with linker **L4**, which features two additional methyl groups at each end of the linker. In both cases of **TFB** and **TFPT**, only the monomeric species, namely **MC-4** and **MC-7**, were observed, with conversion rates of 92% and 86%, respectively. The scenario was quite distinct with the shorter linker **L2**, as its reaction with **TFB** exclusively yielded monomeric cage **MC-2** with a conversion rate of 88%; when reacted with **TFPT**, it only produced catenated cage **DCC-3** with a conversion rate of 63%.

These results can be readily understood by referring to Fig. 4a–c. In the case of the cage formed with **TFB** and **L2**, the size of each window is insufficient to accommodate a twisted linker of another cage (Fig. 4a). Consequently, the formation of the corresponding dimeric catenane is prohibited. In contrast, the windows of the cage formed with **TFPT** and **L2** are large enough to accommodate the three twisted linkers of another cage, thereby facilitating the formation of a dimeric cage-catenane (Fig. 4b). However, when a longer linker **L4** is reacted with **TFPT** to form a monomeric cage, each of its window becomes too narrow to host two twisted linkers required for the formation of a trimeric catenane. Although a single twisted linker for a dimeric catenane can still fit in the window, its formation is entropically unfavoured.

In a word, these findings demonstrate that the enhanced π-π stacking of the panel is crucial in facilitating the catenation of cage monomers. Furthermore, the length of a linker needs to be appropriate, allowing for the hosting of two panels within the cavity of the corresponding monomeric cage. Besides, the window size of the cage needs to be sufficient to accommodate two twisted linkers within each window. These conditions were essential for the formation of a trimeric cage-catenane.

**Confirmation of inherent topological chirality of TCC-1**

With the progress of synthetic techniques in chemical topology, the precise interweaving of molecular subunits has granted access to appealing topological properties[33–36], including topological chirality. Unlike simple covalent stereogenic structures that can be transformed into their enantiomers by adjusting the Euclidean properties of molecular bonding (such as bond lengths and angles) while keeping atomic connectivity intact, stereoisomers of topological chirality remain invariant and cannot be interchanged without breaking and forming new atomic connections[37–40].

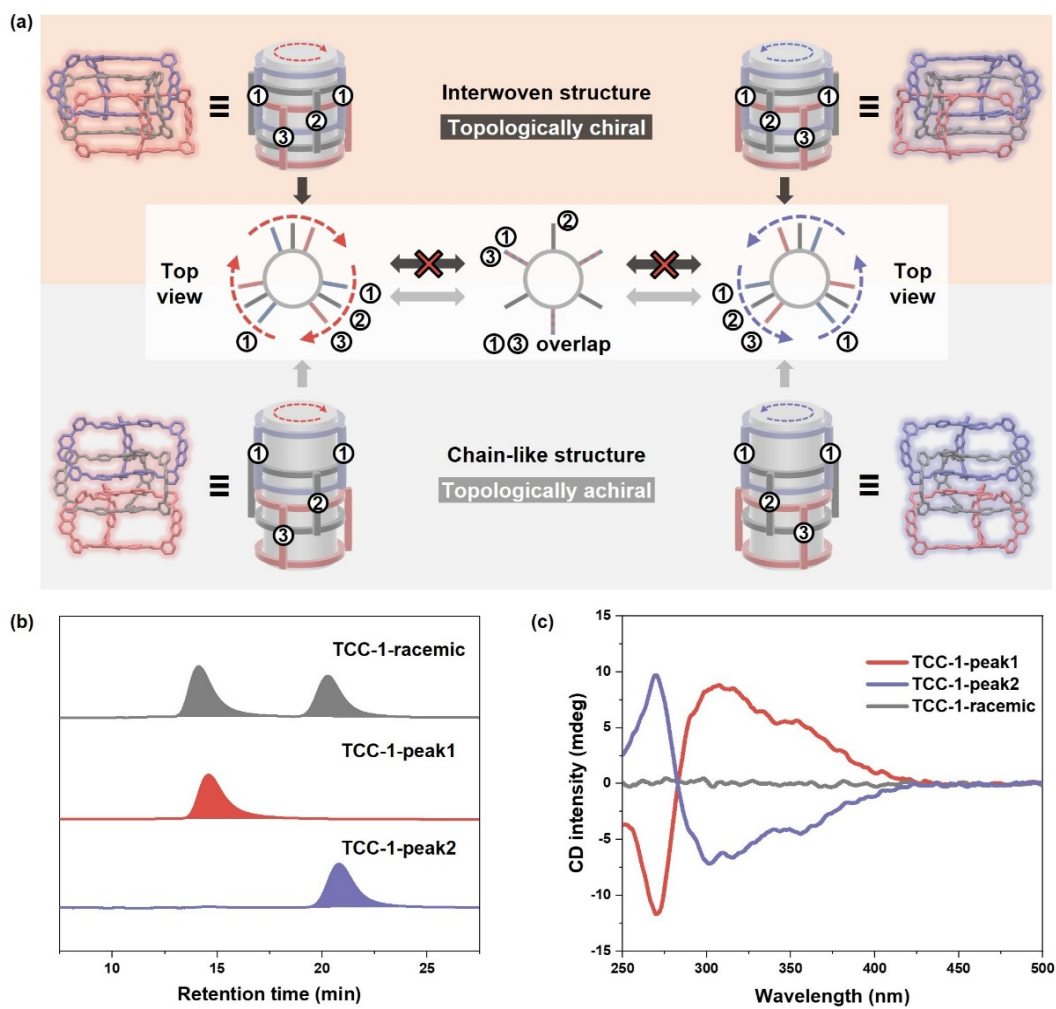

**Fig. 5 A pair of interwoven trimeric cage-catenane enantiomers with topological chirality**. **a**, Schematic demonstration of interwoven **TCC-1** with topological chirality, as the two enantiomers cannot be interconverted without bond cleavage; while its chain-like analogue is topologically achiral, since the two mirror images can be converted by simply rotating the upper and lower cage monomers of the chain-like catenane. The numbers 1–3 correspond to the linkers of three compositional cage monomers in a trimeric cage-catenane, which are used to identify its rotation direction. **b,c**, Chiral-HPLC chromatograms and CD spectra of racemic mixture and homochiral **TCC-1**.

As mentioned above, the mirror images of interwoven trimeric cage-catenane **TCC-1** cannot be converted without bond fission, which therefore should be considered as topologically chiral. The topological chirality of the interwoven **TCC-1** can be unambiguously illustrated using a simplified cartoon model, where the linkers are labelled as **1**–**3** that belong to three monomeric cages (Fig. 5a). The rotation direction of the numbers **1** to **3** is used to determine the handedness of **TCC-1**, which can be either clockwise or counter-clockwise. The transformation from a clockwise structure to a counter-clockwise structure requires an intermediate state where linker **1** overlaps with linker **3**, as shown with the top view. However, the interwoven structure cannot attain this state due to the interpenetration of the cages to which linkers **1** and **3** belong. Consequently, it is impossible to convert an interwoven **TCC-1** molecule into its mirror image simply by changing the conformation of cages without bond cleavage (top view in Fig. **5a**). On the other hand, as the linkers **1** and **3** in the chain-like catenane are not interpenetrated, it can easily reach the intermediate state by rotating the compositional cages. Therefore, the interwoven **TCC-1** molecule exhibits topological chirality, whereas the chain-like structure is topologically achiral.

This inherent topological chirality was confirmed by chiral-HPLC and CD spectroscopy (Fig. **5b,c**). The chiral-HPLC chromatogram of the as-purified **TCC-1** displays a pair of peaks with equal areas, and its CD spectrum shows no optical activity. This racemic mixture was resolved using semi-preparative chiral HPLC, resulting in the isolation of enantiopure **TCC-1** enantiomers, as confirmed by CD spectroscopy. For instance, the CD spectrum of the first peak (in red) with a retention time of *ca*. 15 min in the chiral-HPLC chromatogram (Fig. **5b**) exhibits a positive-to-negative bisignate curve, with a positive cotton effect at 300 nm followed by a negative one at 270 nm (Fig. **5c**), indicating a *P*-conformation. Additionally, the CD spectra of the two resolved enantiomers exhibit mirror-image profiles, underscoring their opposite chirality.

**Conclusions**

Guided by a mathematical analysis showing that the interwoven trimeric catenane is favoured over its chain-like analogue, we rationally designed and synthesized the former by a [9+6] cycloimination of a diamine linker and a trialdehyde aromatic panel. The linker should possess an appropriate length to accommodate two panels within the cavity of the monomeric cage. The enhanced π-π stacking of the aromatic panel plays a crucial role in facilitating the catenation. This advantageous template effect effectively inhibits the formation of the chain-like catenanes, making the interwoven trimeric catenane as the dominating species.

Similar to the interwoven [3]catenane[13–15] that features three interpenetrated rings and is deemed topologically chiral[13,39], our interwoven trimeric cage-catenane, with a novel and more sophisticated embedded graph (Fig. **1a**), also exhibits topological chirality. The chirality in such structure is revealed for the first time, through various characterization techniques including chiral-HPLC, CD spectroscopy, and SC-XRD.

The mechanical bonding[10] of the multi-annulated monomers, showcased here with bicyclic cage compound, would offer a new venue for constructing topologically sophisticated structures, including but not limited to poly[n]catenane of different topology[41–45]. Traces of catenated oligomers have been observed in the current study (Supplementary Fig. S75). This promising preliminary result prompted us to revisit the rational design of precursors and optimisation of the reaction conditions, so that the interwoven catenation can be surprised to favour chain growth of the polycatenanes. The synthesis and property exploration of these novel polymers are ongoing in our laboratories.

## Methods

**Synthesis of interwoven trimeric cage TCC-1**

Panel **TFPT** (197 mg, 0.5 mmol) and linker **L3** (320 mg, 0.75 mmol) were dissolved in chloroform (150 mL), followed by adding catalytic amount of **TFA** (100 μL of chloroform solution with a concentration of 0.75 mol/L) to afford a clear solution. The imine condensation reaction proceeded for 5 hours at 30 ºC with stirring. After the reaction, $NaHCO_3$ and anhydrous $Na_2SO_4$ were added into the mixture and stirred for 5 minutes. The mixture was filtered and concentrated by rotary evaporation. The concentrated solution was dripped into hexane and the precipitate was collected by centrifugation as the crude product. The crude product was purified by recycling GPC with DCM as the mobile phase to yield **TCC-1** as a yellow powder (303 mg, 62%). $^1$H NMR (700 MHz, $CD_2Cl_2$, 298 K), $\delta$ 8.41 (s, 6H), 8.25 (d, $J$ = 7 Hz, 12H), 8.08 (s, 6H), 7.75–7.74 (m, 36H), 7.71 (d, $J$ = 7 Hz, 6H), 7.60 (s, 6H), 7.50–7.42(m, 36H), 7.36 (d, $J$ = 7 Hz, 6H), 7.24–7.18 (m, 30H), 7.14 (t, $J$ = 7 Hz, 6H), 7.09–7.08 (m, 12H), 6.99 (s, 6H), 6.92–6.86 (m, 18H), 6.75 (s, 6H), 4.74–4.78 (m, 6H), 4.54–4.52 (m, 6H), 4.13, (m, 6H), 1.48–1.44 (dd, $J$ = 7 Hz, $J$ = 14 Hz, 54H), 1.25–1.04 (m, 108H). $^{13}$C NMR (175 MHz, $CD_2Cl_2$, 298 K), $\delta$ 170.40, 169.72, 169.45, 160.94, 160.43, 160.39, 154.20, 154.08, 153.79, 153.70, 153.40, 153.29, 140.76, 140.21, 140.04, 139.98, 139.62, 139.31, 138.16, 138.02, 137.49, 135.15, 134.99, 131.33, 131.28, 130.75, 130.70, 130.44, 130.35, 129.30, 128.87, 128.74, 128.61, 128.48, 128.41, 128.22, 127.98, 127.73, 127.52, 124.40, 124.32, 120.86, 120.83, 120.80, 120.55, 120.39, 119.40, 107.26, 107.07, 107.05, 70.62, 70.28, 69.92, 22.24, 22.20, 22.06, 21.98, 21.86, 21.53. MALDI-TOF-MS: calculated for $C_{396}H_{324}N_{36}O_{18}$ $[M+H]^+$ 5875.574; found, 5875.903.

## Acknowledgements

We thank financial support from the Science and Technology Commission of Shanghai Municipality (21JC1401700) and the National Natural Science Foundation of China (T2325017, 21890733, 22071153, 22271187). The crystallographic experiments were conducted with beam line BL17B1 supported by Shanghai Synchrotron Radiation Facility. We also thank Dr. Ruibin Wang (CD) and Dr. Hang Wang (MALDI-TOF MS) at the Instrumental Analysis Centre of SJTU, and Mr. Zhiyong Peng and Prof. Wei Wang at East China Normal University for the assistance in chiral HPLC characterization and seperation.

## Author contributions

L.C. and S.Z. conceived the study. L.C., K.Y. and Z.C. did the experimental work. L.C., Z.C., W.W., C.L. and S.Z. analysed and interpreted the results. Z.C., W.W. and C.C. performed probability calculations and computational studies. S.Z. prepared the manuscript, which was edited by all the authors.

## Competing interests

The authors declare no competing interests.